\documentclass[11pt,twoside]{article}
\usepackage{asp2014}

\aspSuppressVolSlug
\resetcounters

\begin{document}

\title{A possible solution to the Lyman/Balmer line problem in hot DA white dwarfs}
\author{S. P. Preval,$^1$ M. A. Barstow,$^1$, N. R. Badnell$^2$, J. B. Holberg$^3$ and I. Hubeny$^4$
\affil{$^1$Department of Physics and Astronomy, University of Leicester, University Road, Leicester, LE1 7RH, UK; \email{sp267@le.ac.uk},\email{mab@le.ac.uk}}
\affil{$^2$Department of Physics, University of Strathclyde, Glasgow, G4 0NG, UK; \email{badnell@phys.strath.ac.uk}}
\affil{$^3$Lunar and Planetary Laboratory, Sonnett Space Sciences Bld, University of Arizona, Tucson, AZ 85721, USA; \email{holberg@argus.lpl.arizona.edu}}
\affil{$^4$Steward Observatory, University of Arizona, 933 N. Cherry Ave, Tucson, AZ 85721, USA; \email{hubeny@as.arizona.edu}}}

\paperauthor{Simon P. Preval}{sp267@le.ac.uk}{ORCID_Or_Blank}{University of Leicester}{Author1 Department}{Leicester}{State/Province}{LE1 7RH}{UK}
\paperauthor{Martin A. Barstow}{mab@le.ac.uk}{ORCID_Or_Blank}{University of Leicester}{Author2 Department}{Leicester}{State/Province}{LE1 7RH}{UK}
\paperauthor{Nigel R. Badnell}{badnell@phys.strath.ac.uk}{ORCID_Or_Blank}{University of Strathclyde}{Author3 Department}{City}{State/Province}{G4 0NG}{UK}
\paperauthor{Jay B. Holberg}{holberg@argus.lpl.arizona.edu}{ORCID_Or_Blank}{University of Arizona}{Author3 Department}{City}{State/Province}{AZ 85721}{USA}
\paperauthor{Ivan Hubeny}{hubeny@as.arizona.edu}{ORCID_Or_Blank}{University of Arizona}{Author3 Department}{City}{State/Province}{AZ 85721}{USA}

\begin{abstract}
Arguably, the best method for determining the effective temperature ($T_{\mathrm{eff}}$) and surface gravity (log $g$) of a DA white dwarf is by fitting the Hydrogen Lyman and Balmer absorption features. However, as has been shown for white dwarfs with $T_{\mathrm{eff}}$>50,000K, the calculated value from the Lyman and Balmer lines are discrepant, which worsens with increasing temperature. Many different solutions have been suggested, ranging from the input physics used to calculate the models, to interstellar reddening. We will focus on the former, and consider three variables. The first is the atomic data used, namely the number of transitions included in line blanketing treatments and the photoionization cross sections. The second is the stark broadening treatment used to synthesise the Lyman and Balmer line profiles, namely the calculations performed by \cite{lemke97a} and \cite{tremblay09a}. Finally, the third is the atmospheric content. The model grids are calculated with a pure H composition, and a metal polluted composition using the abundances of \cite{preval13a}. We present the preliminary results of our analysis, whereby we have determined the $T_{\mathrm{eff}}$ for a small selection of white dwarfs. We plan to extend our analysis by allowing metallicity to vary in future model grids.
\end{abstract}

\section{Introduction}
An accurate knowledge of $T_{\mathrm{eff}}$ and log $g$ allows us to deduce many aspects of a white dwarf's evolution. Currently, the best way of determining both of these quantities is by fitting a calculated model atmosphere to either the hydrogen Lyman or Balmer line absorption series \citep{holberg86a}. This is not without it's drawbacks, however, as the shape of the Lyman/Balmer absorption features is also dependent upon the metal species present in the atmosphere. Therefore, inaccurate atomic data used to calculate model atmospheres may also result in an inaccurate determination of the $T_{\mathrm{eff}}$ and log $g$. Current calculations of Lyman/Balmer line profiles also utilise pre-calculated Stark broadening tables to reduce calculation time. As with the atomic data, if the broadening tables are not accurate enough, then this will also propagate errors to the determined $T_{\mathrm{eff}}$ and log $g$.

In an analysis by \cite{barstow01a}, the authors measured $T_{\mathrm{eff}}$ and log $g$ for a sample of DA white dwarfs that had data in the Lyman/Balmer line regions from ORFEUS, HUT, and the Far Ultraviolet Spectroscopic Explorer (FUSE) in the UV, and ground based telescopes in the optical. It was shown that the $T_{\mathrm{eff}}$ determined from the Lyman/Balmer line series disgreed for several white dwarfs, however, this was regarded as being due to systematics present in the observational data. An improved study by \cite{barstow03a} considered a larger data set where all UV data originated from FUSE. A uniform pipeline was developed and applied to the UV data in an attempt to reduce any systematics present. The determined Lyman/Balmer $T_{\mathrm{eff}}$ up to $\sim$50,000K was found to be in agreement for both series, however, a statistically significant discrepancy became apparent for stars with $T_{\mathrm{eff}}$ greater than this. This discrepancy has affectionately been dubbed "The Lyman/Balmer line problem".

In this proceeding, we present preliminary results from our study of the Lyman/Balmer line problem in hot DA white dwarfs. We approach the problem from three perspectives. The first is the opacity used to calculate model atmospheres. This involves using atomic data from the \cite{kurucz92a} (Ku92 hereafter) and the \cite{kurucz11a} (Ku11 hereafter) data releases. The Ku92 release contains $\sim{10^6}$ and $\sim{10^8}$ Fe/Ni {\sc iv-vi} transitions respectively. The second is the broadening tables used to synthesise the Lyman/Balmer line profiles. Two of these tables are investigated, one calculated by \cite{lemke97a}, and another from \cite{tremblay09a}. The third is the composition of the atmosphere. We calculate model atmospheres with a pure hydrogen composition, and a metal polluted composition.

\section{Observations}
Seven white dwarfs are chosen that have been observed both in the UV and optical wavebands. All of these bar WD1254+233 (GD153) are thought to be polluted with metals. Two of these stars, namely WD0501+527 (G191-B2B), and WD1254+233, are calibration standards \citep{bohlin04a} and have well measured $T_{\mathrm{eff}}$. The remaining five, WD0229-481, WD0556-375, WD0621-376, WD1738+669, and WD2211-495 have been observed to have $T_{\mathrm{eff}}$ $>50,000$K, and have exhibited the Lyman/Balmer problem. The UV data for each of these stars is taken from FUSE observations. Where possible, FUSE datasets taken using the LWRS aperture are used as flux losses due to source drift are minimised. Stars with multiple datasets are coadded to improve the signal to noise. The full list of datasets used in this analysis will be included in a future publication.

Optical observations are taken from several sources. In the case of WD0501+527 and WD1254+233, coadded G430L observations from the Space Telescope Imaging Spectrometer (STIS) aboard the Hubble Space Telescope are used. For the remaining five stars, objects with declinations $>0^{\circ}$ were observed by the Steward Observatory, while objects with declinations $<0^{\circ}$ were observed by the South African Astronomical Observatory. Full details on the optical observations are given in \cite{marsh97a}.

\section{Model atmosphere calculations}
Model atmospheres used in this study are calculated using {\sc tlusty} \citep{hubeny88a} version 201, and are synthesised with {\sc synspec} version 49 \citep{hubeny11a}. Each grid is calculated with $T_{\mathrm{eff}}$ ranging from 35,000 to 100,000K in steps of 2,500K, and log $g$ from 6.5 to 9.5 in steps of 0.25 dex. It has been shown by \cite{barstow98a} that the inclusion of heavy metals in model atmospheres will affect the determined $T_{\mathrm{eff}}$. This result is investigated in this study by including two atmospheric compositions. The first is a pure hydrogen composition, and the second is a metal polluted composition using the abundances from \cite{preval13a}, with C/H=$1.72\times{10}^{-7}$, N/H=$2.16\times{10}^{-7}$, O/H=$4.12\times{10}^{-7}$, Al/H=$1.60\times{10}^{-7}$, Si/H=$3.68\times{10}^{-7}$, P/H=$1.64\times{10}^{-8}$, S/H=$1.71\times{10}^{-7}$, Fe/H=$1.83\times{10}^{-6}$, and Ni/H=$1.01\times{10}^{-6}$ as number fractions.

\cite{tremblay09a} calculated new Stark broadening tables, including additional souces of opacity not accounted for in \cite{lemke97a}'s tables. As the line shapes of the Lyman/Balmer series are affected by $T_{\mathrm{eff}}$, log $g$, and the broadening treatment, it is imperative to calculate the synthetic profiles as accurately as possible. To assess the effects of using either the \cite{lemke97a} and the \cite{tremblay09a} calculations, all model grids are synthesised using either of these broadening tables in turn.

The completeness of the atomic data supplied to model atmosphere can be inferred to have an effect upon the determined $T_{\mathrm{eff}}$. Radiative levitation calculations by \cite{chayer95a} were found to be dependent upon the number of transitions included in the calculation. We investigate this possibility by calculating the metal polluted model atmospheres using either the Ku92 dataset, or the Ku11 dataset. The difference between these datasets is the number of energy levels and transitions available for the ions Fe/Ni {\sc iv-vi}. Ku92 is supplemented with photoionization (PI) cross sections calculated by the Opacity Project for Fe {\sc iv-vi}, while Ni {\sc iv-vi} is supplemented with PI data calculated using an hydrogenic approximation. The Ku11 data is not accompanied by PI cross section data. We therefore calculated this data using the atomic collision package {\sc autostructure} \citep{badnell86a,badnell97a,badnell11a}. Full details of this calculation will be provided in a future publication.

\section{Results/Discussion}
As expected, the largest Lyman/Balmer line discrepancies occur when pure H grids are used to measure $T_{\mathrm{eff}}$. Curiously, however, in the case of WD1254+233, which is known to be a pure H atmosphere star, the discrepancy between the Lyman/Balmer line $T_{\mathrm{eff}}$ is not affected by the metal content, but is resolved upon using the \cite{lemke97a} broadening tables. In the case of the metal rich stars, the discrepancies between the Lyman/Balmer $T_{\mathrm{eff}}$ are greatly reduced when using a metal polluted model grid, and are improved further when these grids are synthesised using the \cite{tremblay09a} tables. In Figure \ref{fig:teffres}, we have plotted the measured $T_{\mathrm{eff}}$ using the Lyman/Balmer lines for the different types of model grid used. We have also tabulated the $T_{\mathrm{eff}}$ determinations in Table \ref{table:results}. Interestingly, the use of Ku11 data in model atmosphere calculations does not appear to have any noticable effect upon the determined $T_{\mathrm{eff}}$, regardless of the Stark broadening table employed. It is for this reason we have omitted the Ku11 results from Figure \ref{fig:teffres}. The next step to try and improve the agreement between the Lyman/Balmer is to allow variations in the metal abundances. This will be done by calculating a set of model grids for different multiples of the G191-B2B abundances used in this proceeding. We also plan to include more stars in our sample.

\articlefigure{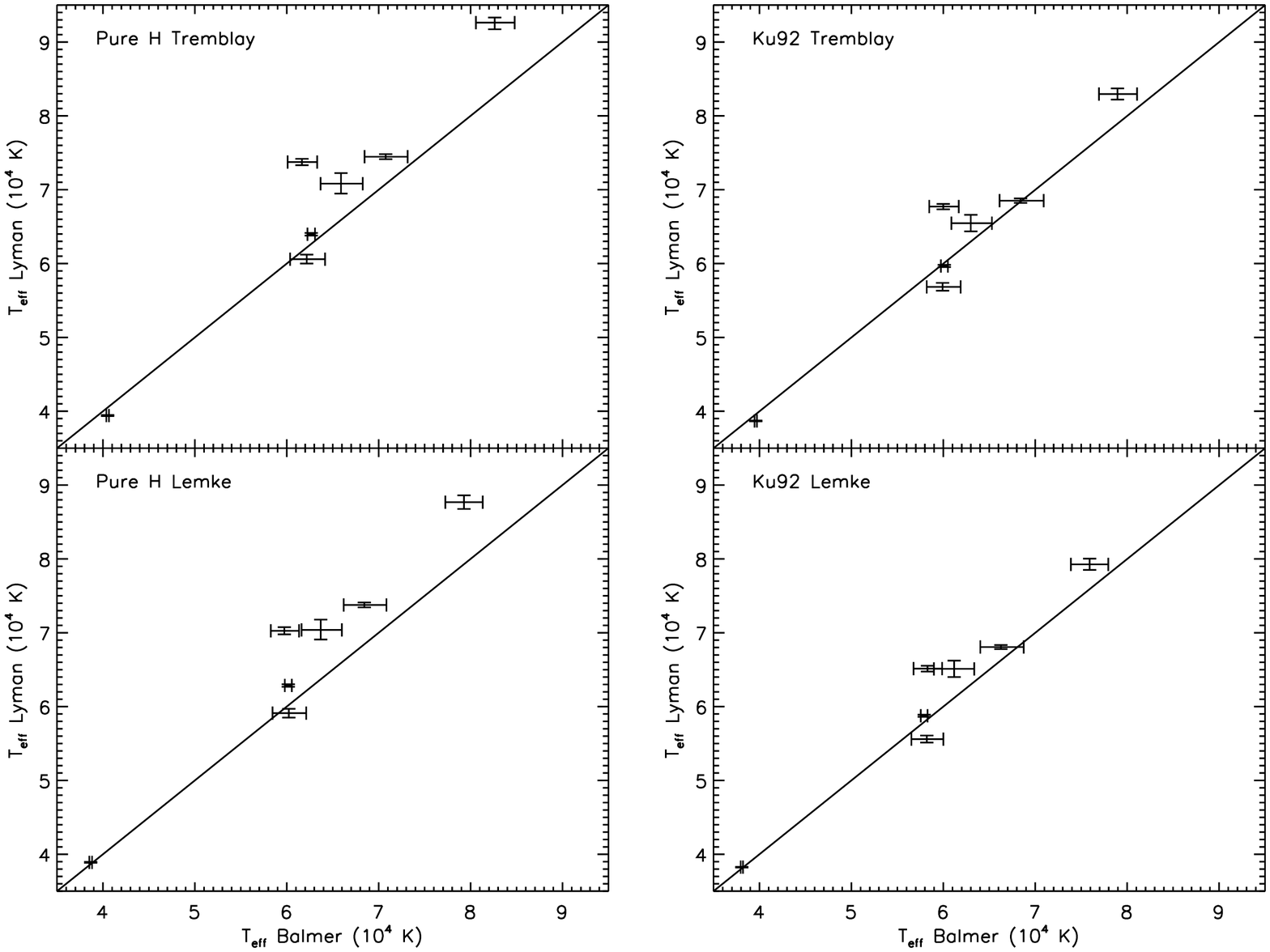}{fig:teffres}{Plot of Lyman $T_{\mathrm{eff}}$ vs Balmer $T_{\mathrm{eff}}$ determined for the seven white dwarfs. The upper two panels are for models synthesised using the Tremblay tables, and the bottom two are for models synthesised with the Lemke tables. The two panels on the left are for pure H atmospheres, and the two panels on the right are for a metal polluted atmosphere using the Ku92 data release with Opacity Project cross sections. The straight line is for equal Lyman/Balmer $T_{\mathrm{eff}}$.}

\begin{table}[!ht]
\caption{$T_{\mathrm{eff}}$ determinations from seven white dwarfs in K. For each white dwarf, the upper row corresponds to Lyman line measurements, while the lower row corresponds to Balmer line measurements.}
\smallskip
\begin{center}
{\small
\begin{tabular}{lllllll}  
\tableline
\noalign{\smallskip}
                 & Lemke &            &           & Tremblay & & \\
\noalign{\smallskip}
WD name & Pure H & Ku92 & Ku11 & Pure H & Ku92 & Ku11 \\
\noalign{\smallskip}
\tableline
\noalign{\smallskip}
WD0229-481  & $59,116_{-599}^{+585}$ & $55,600_{-467}^{+485}$ & $55,670_{-467}^{+487}$             & $60,596_{-594}^{+621}$ & $56,843_{-514}^{+540}$ & $56,914_{-515}^{+540}$ \\
\noalign{\smallskip}
                       & $60,227_{-1779}^{+1885}$ & $58,206_{-1677}^{+1807}$ & $58,143_{-1658}^{+1806}$ & $62,173_{-1807}^{+1995}$ & $59,922_{-1721}^{+1960}$ & $59,929_{-1765}^{+1907}$ \\
\noalign{\smallskip}
WD0501+527 & $62,863_{-168}^{+166}$ & $58,754_{-143}^{+145}$ & $58,769_{-141}^{+143}$             & $63,973_{-177}^{+179}$ & $59,671_{-152}^{+154}$ & $59,693_{-149}^{+153}$ \\
\noalign{\smallskip}
                       & $60,165_{-381}^{+383}$ & $57,919_{-370}^{+366}$ & $57,888_{-369}^{+365}$             & $62,658_{-402}^{+412}$ & $60,113_{-377}^{+390}$ & $60,102_{-375}^{+388}$ \\
\noalign{\smallskip}
WD0556-375  & $70,383_{-1304}^{+1401}$ & $65,114_{-1103}^{+1106}$ & $65,001_{-1072}^{+1099}$ & $70,823_{-1341}^{+1424}$ & $65,463_{-1099}^{+1128}$ & $65,344_{-1081}^{+1123}$ \\
\noalign{\smallskip}
                       & $63,691_{-2083}^{+2304}$ & $61,173_{-2189}^{+2190}$ & $61,127_{-2177}^{+2179}$ & $65,902_{-2217}^{+2359}$ & $62,986_{-2102}^{+2301}$ & $62,956_{-2068}^{+2278}$ \\
\noalign{\smallskip}
WD0621-376  & $70,268_{-481}^{+489}$ & $65,142_{-393}^{+384}$ & $64,988_{-389}^{+380}$            & $73,740_{-426}^{+444}$ & $67,716_{-369}^{+361}$ & $67,601_{-373}^{+352}$ \\
\noalign{\smallskip}
                       & $59,731_{-1470}^{+1610}$ & $58,277_{-1513}^{+1594}$ & $58,285_{-1518}^{+1623}$ & $61,650_{-1560}^{+1650}$ & $59,994_{-1523}^{+1680}$ & $60,036_{-1525}^{+1680}$ \\
\noalign{\smallskip}
WD1254+233  & $38,908_{-56}^{+63}$ & $38,256_{-49}^{+51}$ & $38,332_{-49}^{+53}$                        & $39,418_{-65}^{+66}$ & $38,680_{-54}^{+55}$ & $38,773_{-56}^{+57}$ \\
\noalign{\smallskip}
                       & $38,669_{-140}^{+146}$ & $38,083_{-126}^{+131}$ & $38,151_{-127}^{+133}$             & $40,511_{-142}^{+144}$ & $39,595_{-133}^{+132}$ & $39,686_{-135}^{+134}$ \\
\noalign{\smallskip}
WD1738+669  & $87,689_{-918}^{+917}$ & $79,264_{-748}^{+771}$ & $78,870_{-730}^{+755}$            & $92,624_{-894}^{+700}$ & $82,960_{-748}^{+753}$ & $82,440_{-728}^{+718}$ \\
\noalign{\smallskip}
                       & $79,280_{-2024}^{+2037}$ & $75,923_{-2040}^{+2025}$ & $75,701_{-1956}^{+2075}$ & $82,625_{-2049}^{+2166}$ & $78,956_{-2014}^{+2126}$ & $78,763_{-1985}^{+2093}$ \\
\noalign{\smallskip}
WD2211-495  & $73,767_{-329}^{+334}$ & $68,067_{-280}^{+279}$ & $67,900_{-273}^{+272}$             & $74,470_{-333}^{+345}$ & $68,513_{-295}^{+306}$ & $68,341_{-287}^{+297}$ \\
\noalign{\smallskip}
                       & $68,429_{-2232}^{+2417}$ & $66,720_{-2244}^{+2480}$ & $66,228_{-2245}^{+2468}$ & $70,767_{-2300}^{+2384}$ & $68,408_{-2291}^{+2516}$ & $68,409_{-2330}^{+2418}$ \\
\noalign{\smallskip}
\tableline 
\end{tabular}
}
\label{table:results}
\end{center}
\end{table}

\section{Conclusion}
We have presented a preliminary report on our investigation into the Lyman/Balmer line problem from the perspective of the opacity used to calculate the model atmosphere, the Stark broadening tables used to synthesise the Lyman/Balmer line profiles, and the composition of the atmosphere specified in the model. The choice of atomic data, either Ku92 or Ku11 is inconsequential, and the largest improvement is observed when using the Tremblay tables with metal polluted atmospheres.

\acknowledgements FUSE data was obtained from the Mikulski Archive for Space Telescopes (MAST). SPP and MAB acknowledge the support of the Science and Technology Facilities Council (STFC, UK). JBH and IH acknowledge support from the NASA grant NNG056GC46G.

\end{document}